\documentclass[aps,amsmath,amssymb,superscriptaddress]{revtex4-1}

\usepackage{lineno,hyperref}
\usepackage{color}
\usepackage{graphicx}
\usepackage{xcolor} 
\usepackage{lineno}
\usepackage{enumitem}
\usepackage{multirow}

\modulolinenumbers[5]

\begin{document}

\title{Fissile material detection using neutron time-correlations from photofission}

\author{R. A. Soltz}
\email{soltz1@llnl.gov}
\affiliation{Lawrence Livermore National Laboratory, Livermore, CA 94550, USA}
\author{A. Danagoulian}
\affiliation{Passport Systems Inc., North Billerica, MA 01862, USA}
\affiliation{Massachusetts Institute of Technology, Cambridge, MA 02139, USA}
\author{E.~P.~Hartouni}
\affiliation{Lawrence Livermore National Laboratory, Livermore, CA 94550, USA}
\author{M.~S.~Johnson}
\affiliation{Lawrence Livermore National Laboratory, Livermore, CA 94550, USA}
\author{S.~A.~Sheets}
\affiliation{Lawrence Livermore National Laboratory, Livermore, CA 94550, USA}
\author{A.~Glenn}
\affiliation{Lawrence Livermore National Laboratory, Livermore, CA 94550, USA}
\author{S.~E.~Korbly}
\affiliation{Passport Systems Inc., North Billerica, MA 01862, USA}
\author{R.~J.~Ledoux}
\affiliation{Passport Systems Inc., North Billerica, MA 01862, USA}

%
%
%

\date{\today}


\begin{abstract}
The detection of special nuclear materials (SNM) in commercial cargoes is a major objective in the field of nuclear security.  In this work we investigate the use of two-neutron time-correlations from photo-fission using the Prompt Neutrons from Photofission (PNPF) detectors in Passport Systems Inc.'s (PSI) Shielded Nuclear Alarm Resolution (SNAR) platform for the purpose of detecting $\sim$5~kg quantities of fissionable materials in seconds.  The goal of this effort was to extend the secondary scan mode of this system to differentiate fissile materials, such as highly enriched uranium, from fissionable materials, such as low enriched and depleted uranium (LEU and DU).  Experiments were performed using a variety of material samples, and data were analyzed using the variance-over-mean technique referred to as  $Y_{2F}$ or Feynman-$\alpha$.  Results were compared to computational models to improve our ability to predict system performance for distinguishing fissile materials.  Simulations 
were then combined with empirical formulas to generate receiver operating characteristics (ROC) curves for a variety of shielding scenarios. We show that a 10 second screening with a 200~$\mu$A 9~MeV X-ray beam is sufficient to differentiate kilogram quantities of HEU from DU in various shielding scenarios in a standard cargo container. 
\end{abstract}


\maketitle

\section{Introduction}

Fissile materials refer to materials which, due to their nuclear structure, allow for sustained fission chains.  The two most common isotopes which form the basis of fissile materials are $^{235}$U and $^{239}$Pu. Methods for detecting the presence of fissile materials support the goals of national and international programs for nuclear-nonproliferation. An extensive literature exists on potential identification schemes using active interrogation including neutron probes and delayed and prompt neutron signals~\cite{Campbell1978, Hollas2001, Slaughter2003, Lia2004,Dietrich2005}, photon probes and delayed neutron signals~\cite{Jones2000, Jones2002a, Jones2002b, Jones2003a, Jones2005b, Jones2005c, Jones2006}, photon and neutron probes and delayed neutron signals~\cite{Moss2004, Myers2004, Myers2005}, and photon probes and fission product radiation~\cite{Jones2005a, Jones2007}. 
A general review of the various concepts can be found in Ref.~\cite{Runkle}. All of these methods use single-particle signals. Multi-particle schemes, usually two- and three-neutron signals have also been pursued, with photon~\cite{Moss2003, Chichester2012} and neutron~\cite{Deighton1979, Arnone1989,Hollas2005,Nakae2009,McConchie2009a} beams as well as in passive interrogation~\cite{Nixon1982,Krick1984,Cifarelli1985,Dytlewski1993,McConchie2009b,McConchie2009c,Pena2010}, where ambient radiations are used to induce the signals.  
Here we report on an active interrogation method using 9 MeV Bremsstrahlung photons to induce the emission of time-correlated, prompt neutrons measured within the Prompt Neutrons from Photofission (PNPF) system developed at Passport Systems Inc. (PSI)~\cite{pnpf,PNPF_patent,pnpf-short}.

\section{Time-correlation of fission-chain neutrons}
A fission chain reaction occurs in fissile material when the neutrons from each fission diffuse through the material and induce subsequent fissions in other nuclei.  The neutrons emitted from this process are highly correlated, resulting in neutron count distributions that deviate significantly from Poisson distributions produced by random neutron events.

The theory of fission-chain correlations was initially developed by Richard Feynman while at Los Alamos~\cite{Feynman1956}. The aim of that research was to describe the neutron fluctuations in a reactor pile where the measured neutrons originate from fission chains and random decays.  To measure deviations from the Poisson expectation, Feynman defined a normalized second moment,
\begin{equation}
Y_{2F} = \frac{\langle{n^2}\rangle-{\langle n\rangle}^2 - \langle n \rangle}{2\langle n \rangle} \equiv \frac{\rm variance - mean}{2 * \rm mean}
\label{y2f_exp}
\end{equation}
where $n$ is the measured neutron count per unit time, and $\langle n \rangle$ and $\langle n^2 \rangle$ represent averages over a series of time-gates.  Subtracting the mean from the variance enforces $Y_{2F}=0$ for the special case of a Poisson distribution, and division by the mean ensures that the quantity is independent of rate. The subscript ``2F'', not used in the original paper, has been added by others to credit Feynman and to denote that this quantity is derived from the second moment.

This statistic has been used extensively in the study of nuclear reactor cores~\cite{Degweker1997a,Degweker1997b,Degweker1997c, Degweker1999,Degweker2000}, and it has been extended for time varying sources of the fluctuations, e.g. in accelerator driven cores~\cite{Munoz-Cobo1987a, Munoz-Cobo1987b, Degweker1994,Degweker2003,Ballester2005a,Ballester2005b,Ballester2005c}. In these systems the number of neutrons generated spontaneously from radioactive decay greatly exceeds those generated by the introduced beam. 

The idea for using the time-correlation for non-destructive inspection of objects containing fissile material has been pursued for ``passive'' interrogation schemes~\cite{Deighton1979,Nixon1982,Robba1983,Krick1984,Dowdy1985,Cifarelli1985,Arnone1989,Dytlewski1993,Estes1995,McConchie2009b,Nakae2009,Croft2012} (where the natural background initiates the fission chains) and for ``active'' interrogation schemes~\cite{Hollas1999,Moss2003,Hollas2005,McConchie2009a,Chichester2012} (where a beam of particles induces the fission chains).

For neutron-induced fission-chains from an idealized point source, a time-dependent expression for $Y_{2F}$ has been developed by Snyderman and Prasad~\cite{Snyderman2005,Prasad2012,Walston2011},
\begin{eqnarray}
Y_{2F} & = & \left( 1-\frac{1-e^{-\lambda T}}{\lambda T} \right) \epsilon \left[1 + \frac{(M-1)(\nu-1)}{\nu} \right] \frac{\nu_2} {\nu} (M-1).
\label{eq:y2f_induced}
\end{eqnarray}
where,\\
\bigskip
\begin{tabular}{rcl}
$\lambda$ & = & inverse of the neutron correlation time, a convolution of fission-chain and neutron transit times,\\
$T$ & = & time-gate during which neutrons are counted,\\
$\epsilon$ & = & neutron detection efficiency,\\
$M$ & = & neutron multiplication, $\frac{1}{1-k}$, where $k=p\nu$ is k-effective and $p$ is the neutron-induced fission probability,\\
$\nu$ & = & the first moment, equal to the mean number for the neutron distribution from a single induced fission,\\
$\nu_2$ & = & the second combinatorial moment, half the variance for the neutron distribution from a single fission.
\end{tabular}

Eq.~\ref{eq:y2f_induced} illustrates utility of $Y_{2F}$, whcih increases with the square of the multiplication and is therefore sensitive to the enrichment if the geometry of the object and efficiency of the detector are known.  One can construct higher order moments (i.e. $Y_{3F,4F}$) that depend on higher orders of the multiplication, but a corresponding dependence on higher orders of the efficiency makes these statistics more suitable for large acceptance detectors and/or long integration times due to the reduced coincidence rates.  It is also important to note that Eq.~\ref{eq:y2f_induced} is appropriate for neutron beams.  In Sec.~\ref{sec:results} we will modify this expression for photon-induced fission-chains in extended objects.



\section{Experimental Setup}

To study the feasibility of using photo-induced fissions chains to identify and characterize fissile materials we use a subset of the components of the
PSI SNAR platform. The photon beam is generated from electron bremsstrahlung on a water-cooled radiator. The electron current is adjustable and uniform with high duty-cycle. For all experiments a 9 MeV electron kinetic energy was used.  The beam currents are adjustable from 100 to 500 $\mu$Amps.  For the results presented here, the beam current was set to 200~$\mu$Amps to avoid the non-Poisson effects of the data acquisition (DAQ) dead time losses.  This corresponds to a rate of $2 \times 10^{12}$~Hz for photons with energies between 2 and 9~MeV, incident on an approximately 10 $\times$ 10~cm square spot size at the target location.  The photon beam energy spectrum for the 9~MeV Bremsstrahlung radiator is shown in Fig.~1.
%
%
\begin{figure}[htbp]
\begin{center}
\includegraphics[width=0.9\textwidth]{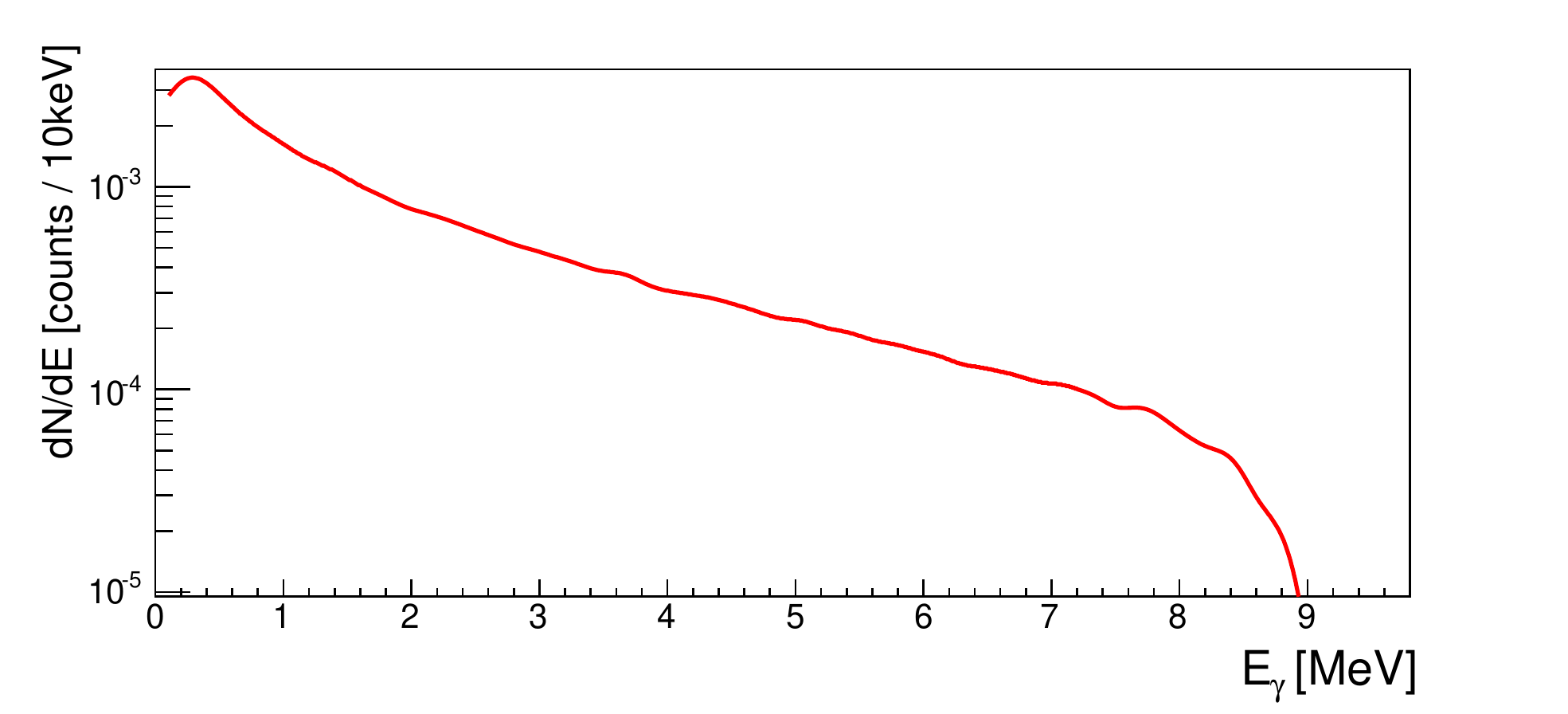}
\label{fig_spectrum}
\caption{Photon energy spectrum in PSI's SNAR system, calculated with a GEANT4 simulation.}
\end{center}
\end{figure}

\begin{figure}[h]
\begin{center}
\includegraphics[width=0.8\textwidth]{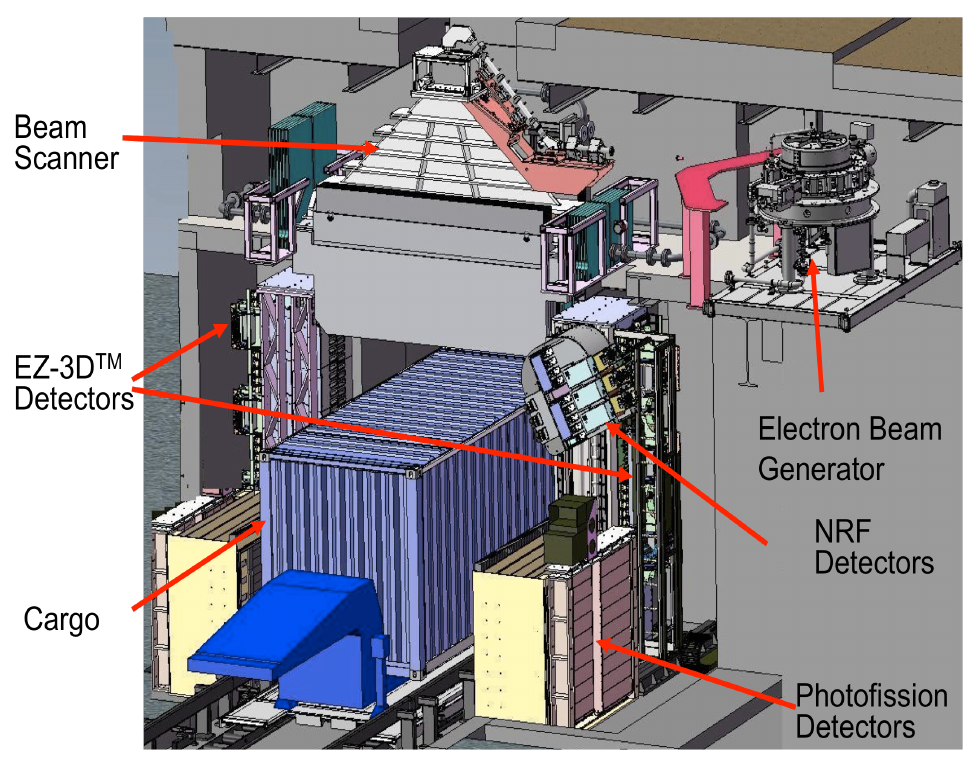}
\caption{Configuration for Passport Systems Inc~Shielded Nuclear Alarm Resolution (SNAR) platform. This analysis uses the photofission detectors shown on the sides.} \label{fig2_passport_snar}
\end{center}
\end{figure}

The full SNAR configuration is shown in Fig.~\ref{fig2_passport_snar}.  The PNPF detectors consist of EJ-309 5~inch diameter liquid scintillator detectors coupled to 5" Hamamatsu photomutilpier tubes arranged in two sets of $2 \times 8$ arrays placed on opposite sides of the cargo container.  
The PNPF arrays have $\sim$5~cm thick high density polyethylene (HDPE) inserts placed in between the detectors to reduce adjacent detector cross-talk that can lead to an artificial $Y_{2F}$ signal.  Before and after tests show that these inserts reduce the cross-talk component by a factor of four. 

\begin{table}
\begin{center}
\begin{tabular}{|r|c|c|c|c|}
\hline
Material & Dimension & Mass & Mult. & Run Duration \\
\hline \hline
Be block  & 7.3 x 7.3 x 10.4 cm & 0.988 kg & 1.000 & 8360 s \\ \hline
DU 1-disc    & 0.2 x 5 cm       & 0.3 kg & 1.015 & 6690 s \\ \hline
DU 2-disc    & 0.4 x 5 cm       & 0.6 kg & 1.026 & 6750 s \\ \hline
DU 3-disc    & 0.6 x 5 cm       & 0.9 kg & 1.035 & 7080 s \\ \hline
HEU 1-disc  & 0.2 x 5 cm       & 0.3 kg & 1.061 & 7080 s \\ \hline
HEU 2-disc  & 0.4 x 5 cm       & 0.6 kg & 1.109 & 7080 s \\ \hline
HEU 3-disc  & 0.6 x 5 cm       & 0.9 kg & 1.155 & 6960 s \\ \hline
HEU+LEU   & 0.2 x 5 cm             &                   &             &            \\
1-stack        & 10 x 10 x 2 cm      & 3.988 kg    &   1.177 & 6410 s \\ \hline
HEU+LEU   & 2 (0.2 x 5 cm)       &                   &             &             \\
2-stack        & 2 (10 x 10 x 2 cm) & 7.976 kg   &  1.304  & 7240 s \\ \hline
HEU+LEU   & 3 (0.2 x 5 cm)       &                   &             &             \\
3-stack        & 3 (10 x 10 x 2 cm) & 11.964 kg  &  1.401  & 6920 s \\ \hline
HEU+LEU   & 3 (0.2 x 5 cm)       &                   &             &             \\
3-stack        & 3 (10 x 10 x 2 cm) & 11.964 kg  &  1.430 & 7080 s \\
+HDPE       & 10 x 10 x 1.27 cm &                   &            &             \\ \hline
HEU+LEU   & 3 (0.2 x 5 cm)       &                   &            &             \\
3-stack        & 3 (10 x 10 x 2 cm) & 11.964 kg  & 1.468 & 6760 s\\
+HDPE       & 10 x 10 x 2.54 cm &                   &            & \\ \hline
HEU+LEU   & 3 (0.2 x 5 cm)       &                   &            & \\
3-stack        & 3 (10 x 10 x 2 cm) & 11.964 kg  & 1.501 & 5980 s \\
+HDPE       & 10 x 10 x 3.81 cm &                   &            & \\ \hline
HEU+LEU   & 3 (0.2 x 5 cm)       &                   &            & \\
3-stack        & 3 (10 x 10 x 2 cm) & 11.964 kg  &  1.527 & 7080 s\\
+HDPE       & 10 x 10 x 5.08 cm  &                  &            & \\ \hline
\end{tabular}
\end{center}
\caption{List of objects exposed to the 9~MeV Bremsstrahlung beam within the Passport Systems SNAR facility.}
\label{tab_objects}
\end{table}

\begin{figure}[h]
\begin{center}
\includegraphics[height=1.5in]{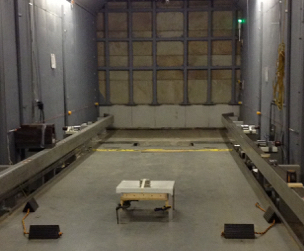}
\includegraphics[height=1.5in]{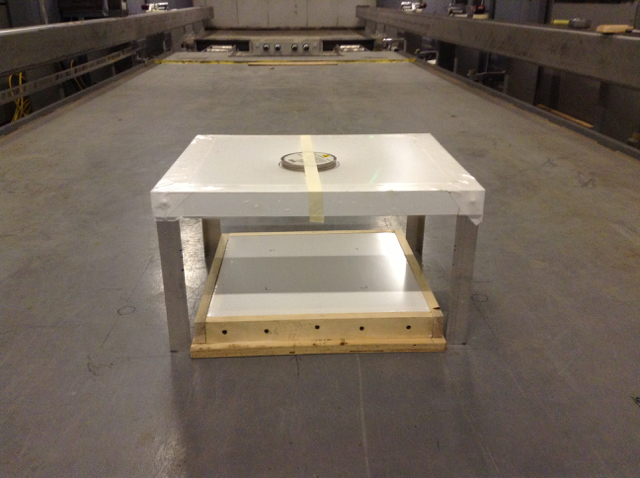}
\caption{Sample object position for SNAR system data collection.}
\label{fig3_object_position}
\end{center}
\end{figure}

Measurements of $Y_{2F}$ were made for a set of objects designed to study the relationship between $Y_{2F}$ and neutron multiplication.  These objects, listed in Table~\ref{tab_objects}, were constructed from discs of depleted uranium (DU) and highly enriched uranium (HEU), and blocks of low enriched uranium (LEU).  The objects were arranged in geometries to achieve a higher range of multiplication values.  Fig.~\ref{fig3_object_position} shows the stacking arrangement for the three HEU discs.
The highest multiplication values were achieved with an interleaved stack of LEU and HEU with HDPE moderators placed above and below the stack to provide neutron reflection. Multiplication values were calculated using MCNP 6.2~\cite{Werner2018} run in the kcode mode.  The control benchmark was obtained with a beryllium block, in order to generate a large quantity of uncorrelated photo-neutrons anticipated to have no measurable $Y_{2F}$ value.  Data were collected from a series of 600-second exposures, which were then combined for each object during the data analysis.  Approximately ten exposures were collected for each uranium configuration, with additional exposures taken for the beryllium object.

\section{Data Analysis}

\subsection{Neutron Identification}

\begin{figure}[htbp]
\includegraphics{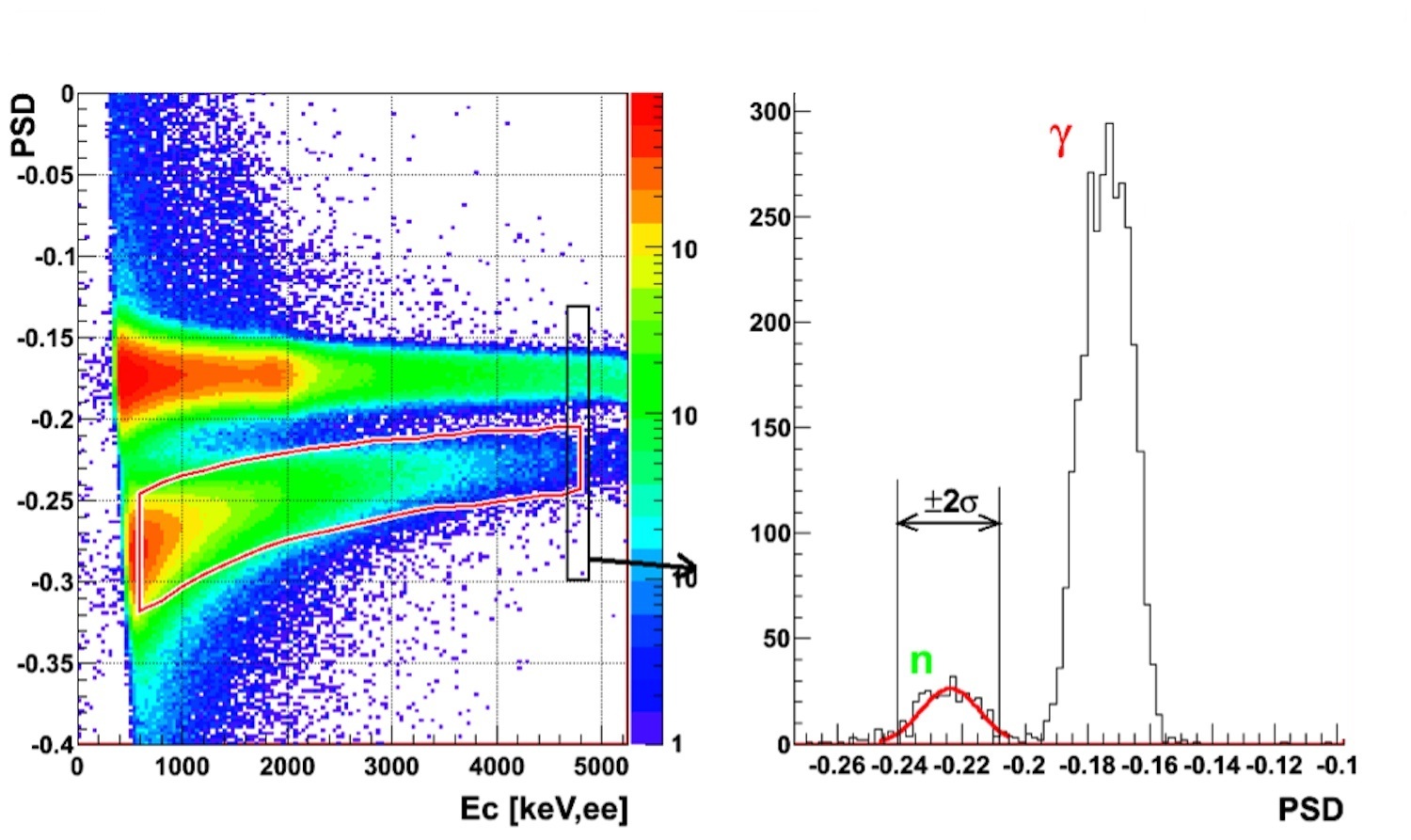}
\caption{(Left) pulse shape descriminiation (PSD) metric vs. electron equivalent keV. The red box indicates the region of signals identified as neutrons.  (Right) separation between neutrons and photons in the region just below 5~MeV. }
\label{fig4_psd}
\end{figure}

The main component of the detection system consists of an array of EJ-309 scintillators, coupled to Hamamatsu photomultiplier tubes (PMT).  The choice of the PMTs was based on their $<3$~ns rise time, thus retaining the sensitivity to the difference between the two main fast components of the light production from the triplet-triplet annihilation.  The increase in triplet-triplet annihilation from the much larger ionization density of the neutron-recoiled proton tracks results in a distinct delay in light output.  Neutrons can be distinguished from photons by their relative fractions of fast and slow light output using various Pulse Shape Descrimination (PSD) techniques, and their deposited energies can be quantified.  The left panel of Fig.~\ref{fig4_psd} shows a histogram of PSD vs. deposited energy, in electron equivalents, for a set of detector events.  The distribution is then split into individual energy bins, and the PSD distribution for every energy bin is fitted with two Gaussians, as can be seen in the right panel of Fig.~\ref{fig4_psd}, which shows the 2D cut used to separate neutrons from photons.  However, this is not sufficient to suppress the very large photon population, created by the intense 200~$\mu $A photon beam.  Therefore, more sophisticated pulse shape discrimination algorithms were developed to reject the photon pileups, which otherwise would contribute a significant background to the neutron population~\cite{pnpf-short}. Overall, the combined pulse shape analysis allows a suppression of the photon population by a factor of $\times10^7$, limiting the beam-on background count to only the cosmogenic neutrons.

\subsection{Multiplicity analysis of neutron events}

\begin{figure}[!h]
\begin{center}
\includegraphics[width=0.95\textwidth]{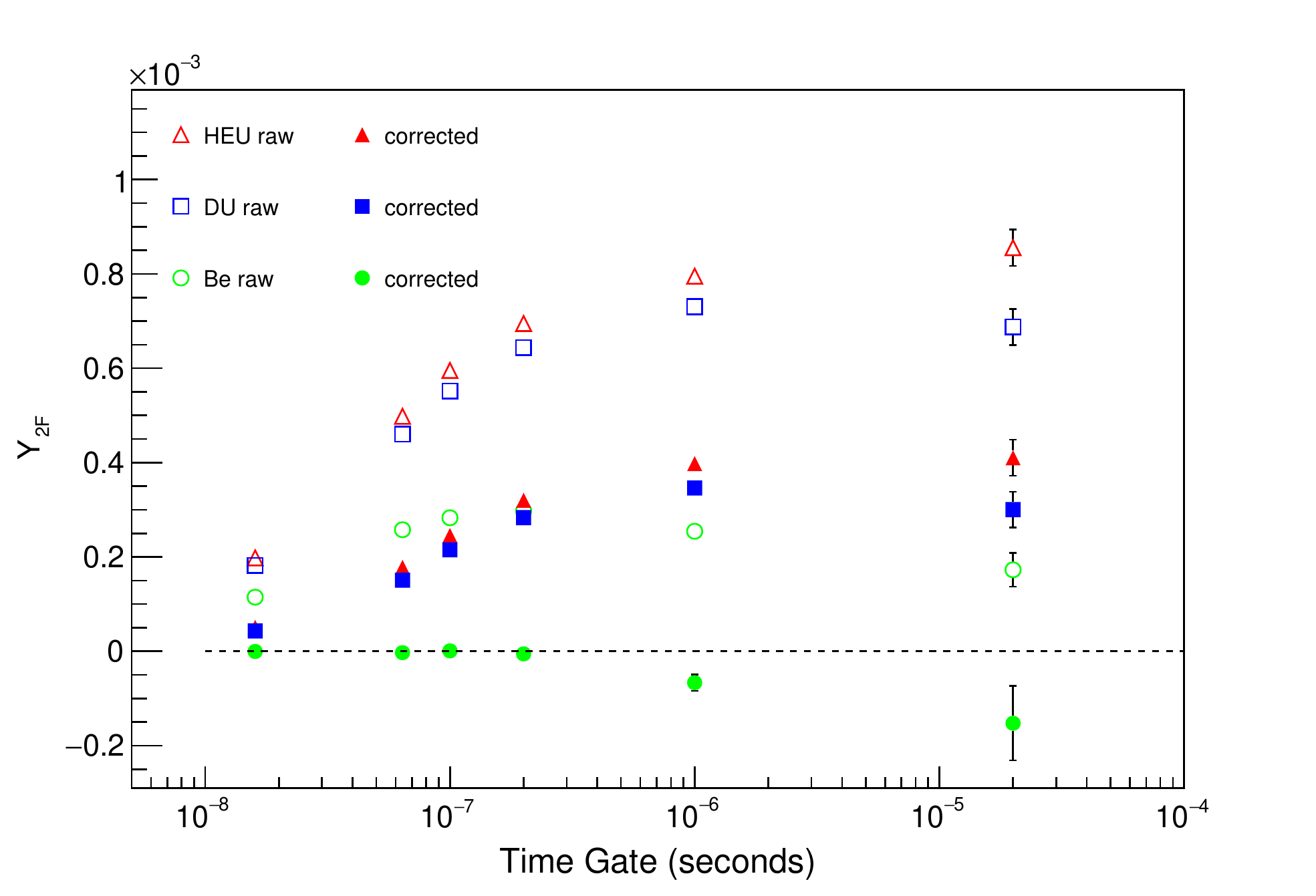}
\end{center}
\caption{The time-gate dependence of the measured $Y_{2F}$ signals for the beryllium, DU, and HEU 3-disc stacks.  Uncorrected values are plotted as open symbols.  The solid symbols show values that have been corrected for cross-talk as explained in Section~\ref{sec:cross_talk}.}
\label{fig_y2f_vstgate}
\end{figure}

The experimental data were collected in a series of 600-second exposures, with some exposures ending prematurely due to disruptions within the data acquisition system due to the extended run times and need for data synchronization.  The identified neutrons in each run were arranged in time-ordered sequence to facilitate binning in time-gates of varying duration.  For clarity, we rewrite Eq.~\ref{y2f_exp} above with explicit averages, where sums are taken over sequential time-gates, $n$ refers to the bin-number corresponding to the neutrons detected within a specific time-gate, and 
$b_n$ is the normalized probability for detecting $n$ neutrons.
\begin{equation}
Y_{2F} = \frac{\sum_n{n^2 b_n} - (\sum_n{n b_n})^2 - \sum_n{n b_n}}{2 \sum_n{n b_n}}.
\label{y2f_bn}
\end{equation}
The errors are calculated using the full covariance matrix for a multinomial distribution~\cite{Barnes2009},
\begin{eqnarray}
\sigma_{Y_{2F}}^2 & = & \frac{1}{N} \sum_{n} b_n (1-b_n) \frac{n^2}{4 \langle n \rangle^4}
\left( n \langle n \rangle - \langle n^2 \rangle - \langle n \rangle^2 \right)^2 \nonumber \\
& & - \frac{1}{N} \sum_{n\neq m} b_n b_m \frac{n m}{4 \langle n \rangle^4}
\left( n \langle n \rangle - \langle n^2 \rangle - \langle n \rangle^2 \right)
\left( m \langle n \rangle - \langle n^2 \rangle - \langle n \rangle^2 \right).
\label{eq:y2f_err}
\end{eqnarray}
The $Y_{2F}$ values and errors for each exposure were calculated separately and checked for outliers before combining the exposures to calculate a final value and error for each object.  The $Y_{2F}$ measurements are listed in Table~\ref{tab_y2f_vstgate} and plotted in Fig.~\ref{fig_y2f_vstgate} along with a set of corrected $Y_{2F}$ values that will be discussed in the next section.  We note that the uncorrected values illustrate the time dependence expected from Eq.~\ref{eq:y2f_induced}.  One can also see from the uncorrected beryllium measurements that a significant non-fission component contributes to the uncorrected $Y_{2F}$ values.
\begin{table}[!h]
\centering
\begin{tabular}{|r | r r | r r  | r r | c |}
\hline
gate  & \multicolumn{2}{|c|}{Be} &  \multicolumn{2}{|c|}{DU} &  \multicolumn{2}{|c|}{HEU} & diff\\
(ns)     &  $Y_{2F}$ & error &   $Y_{2F}$  & error &  $Y_{2F}$  & error & $\sigma$  \\
\hline \hline
16 & 1.15E-4 & 1.70E-6 & 1.82E-4 & 4.08E-6 & 1.98E-4 & 3.46E-6 & \\
corr & -3.25E-7 & 2.23E-6 & 4.33E-5 & 2.19E-6 & 5.05E-5 & 1.98E-6 & 2.4 \\

\hline
64 & 2.58E-4 & 2.88E-6 & 4.60E-4 & 6.62E-6 & 4.98E-4 & 5.63E-6 & \\
corr & -2.85E-6 & 4.44E-6 & 1.51E-4 & 4.16E-6 & 1.78E-4 & 3.79E-6 & 4.8 \\
\hline
100 & 2.83E-4 & 3.31E-6 & 5.52E-4 & 7.35E-6 & 5.95E-4 & 6.30E-6 & \\
corr & 9.33E-7 & 5.59E-6 & 2.15E-4 & 5.04E-6 & 2.45E-4 & 4.55E-6 & 4.5 \\
\hline
200 & 2.98E-4 & 4.19E-6 & 6.44E-4 & 8.32E-6 & 6.95E-4 & 7.24E-6 & \\
corr & -5.42E-6 & 7.87E-6 & 2.83E-4 & 6.20E-6 & 3.20E-4 & 5.67E-6 & 4.4 \\
\hline 
1000 & 2.54E-4 & 8.20E-6 & 7.31E-4 & 1.15E-5 & 7.95E-4 & 1.07E-5 & \\
corr & -6.65E-5 & 1.75E-5 & 3.46E-4 & 1.00E-5 & 3.98E-4 & 9.71E-6 & 3.7 \\
\hline
20000 &  1.73E-4 & 3.56E-6 & 6.87E-4 & 3.85E-5 & 8.56E-4 & 3.85E-5 & \\
corr & -1.53E-4 & 7.89E-5 & 3.00E-4 & 3.80E-5 & 4.10E-4 & 3.83E-5 & 2.0 \\
\hline
\end{tabular}
\caption{The $Y_{2F}$ results from the beryllium, and 3-plate DU and HEU assemblies.  The second row for each time gate, labeled ``corr'' gives the value of $Y_{2F}$ after the cross-talk correction is applied.}
\label{tab_y2f_vstgate}
\end{table}

\subsection{Cross-talk Cuts and Corrections}
\label{sec:cross_talk}

\begin{figure}[htbp]
\begin{center}
\includegraphics[width=0.98\textwidth]{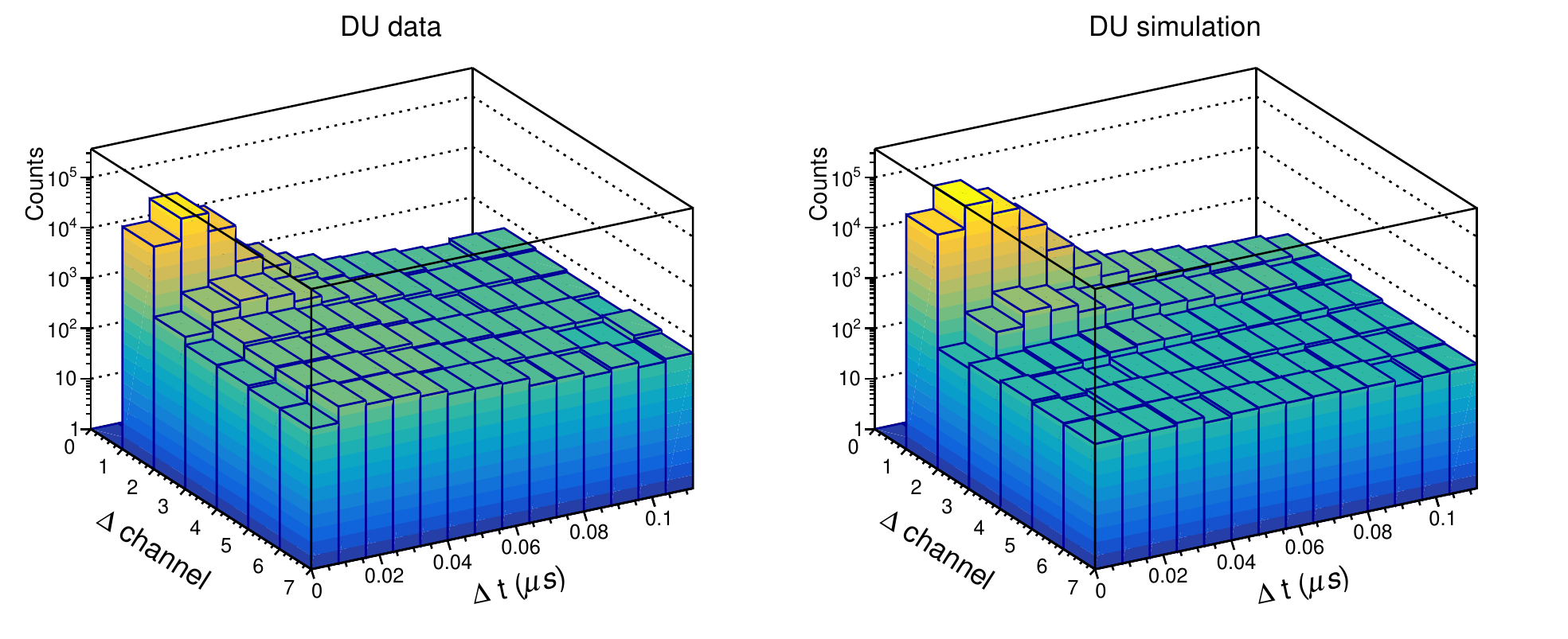}
\caption{(Left) distribution of detected neutron pairs over coincidence time and channel difference for one run of DU data and (right) a corresponding GEANT4 simulation.  Adjacent channels in the same detector-array  column are separated by a distance of 18.5~cm.}
\label{fig6_crosstalk_comparison}
\end{center}
\end{figure}

Neutrons that deposit energy in more than one detector will produce a correlated signal that contributes to the $Y_{2F}$ term.  This effect is shown clearly in Fig.~\ref{fig6_crosstalk_comparison}.
The left panel shows the distribution of pairs of neutron hits for one DU run as a function of the coincidence time and channel separation for channels in the same array column.  Adjacent channels within a single column are separated by a distance of 18.5~cm.  For $\Delta {\rm chan}=1$, 90\% of these channels are separated by this distance, with the remaining 10\% separated by a distance of 2~meters or greater.  $\Delta {\rm chan}=2$, 80\% are separated by a distance of 37~cm.  The corresponding distribution for a GEANT4~\cite{geant4-short} simulation is shown on the the right.  The DAQ trigger logic rejects new signals within 256~ns after a given trigger, leaving zero entries $\Delta {\rm chan}=0$ row of Fig.~\ref{fig6_crosstalk_comparison} creating the empty bins in the top row that is obscured from view.  A significant cross-talk effect is evident in the vertically adjacent and next-nearest neighbor channels ($\Delta {\rm chan}=1,2$).
These enhancements are also visible in the GEANT4 simulation shown in the right panel.  Additional time-dependent structure in the data is attributed to the neutron-correlations, which were not included in this simulation.

\begin{figure}[htbp]
\begin{center}
\includegraphics[width=0.49\textwidth]{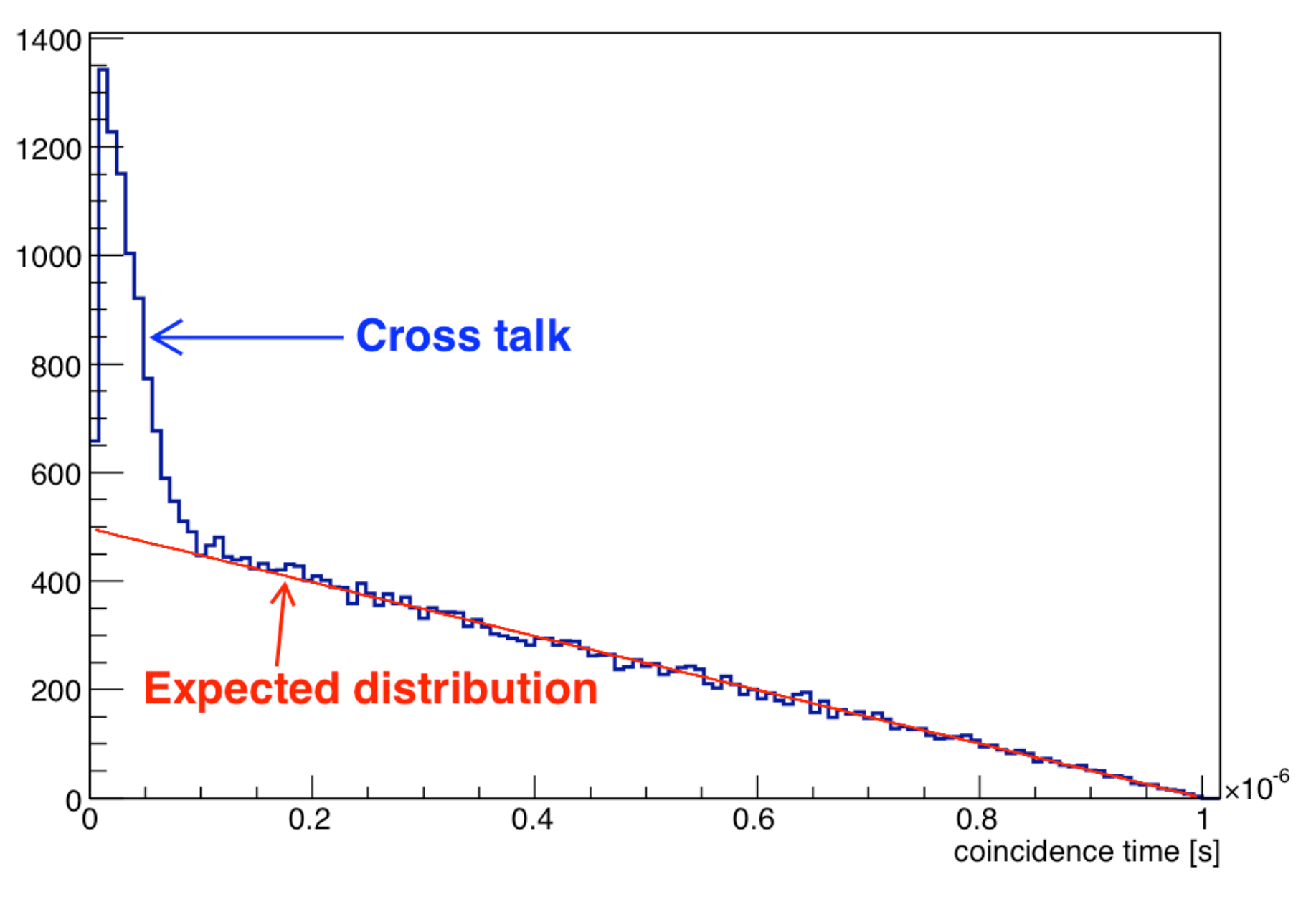}
\includegraphics[width=0.49\textwidth]{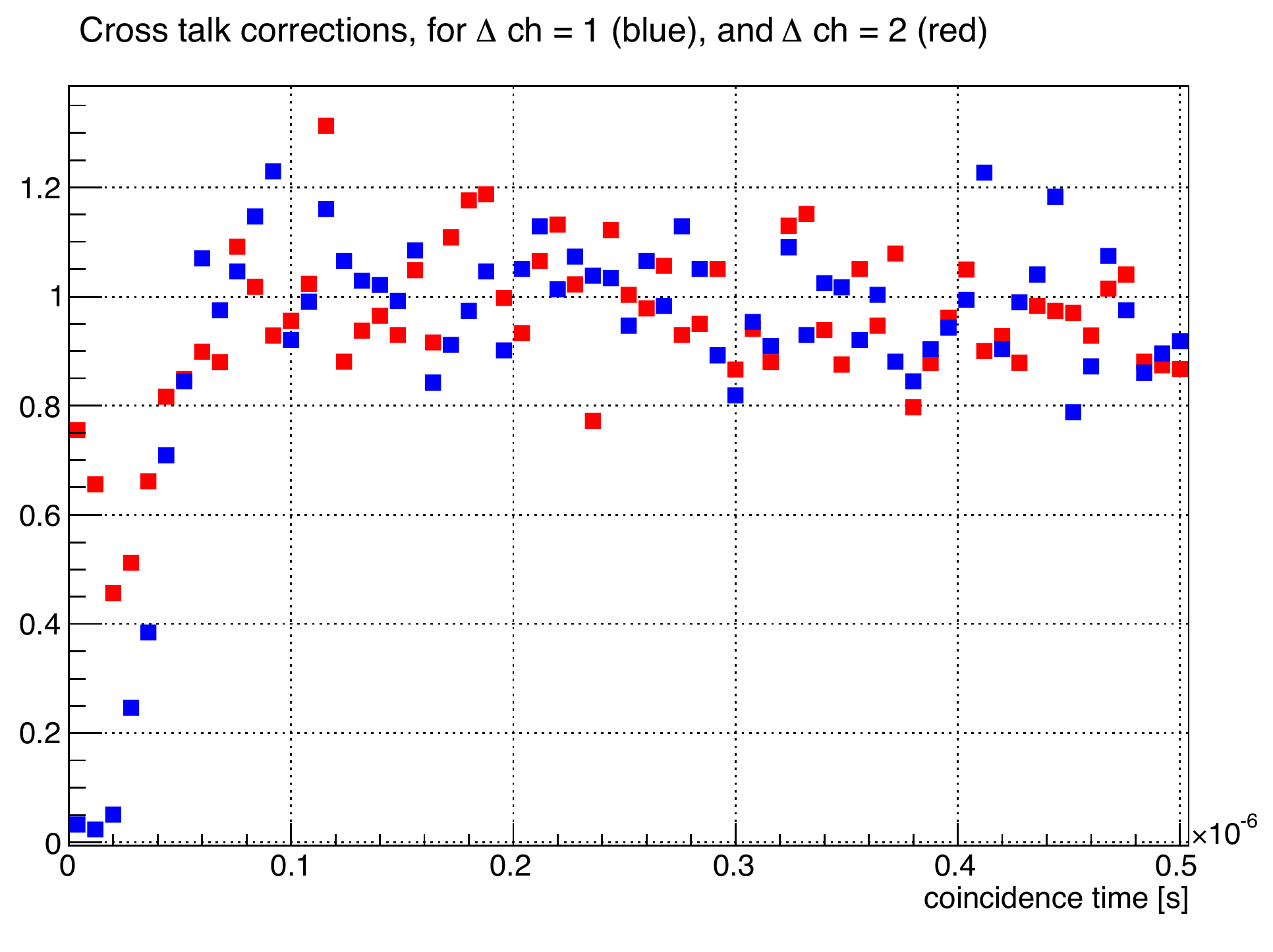}
\caption{(Left) the coincidence time distribution for any two pairs of events within a 1~$\mu$s time gate, for $\Delta ch=1$.  The peak at lower values corresponds to cross-talk between neighboring channels.  (Right) cross-talk correction factors for $\Delta ch=1,2$.} 
\label{fig_dt}
\end{center}
\end{figure}

We have corrected for this cross-talk effect on adjacent channels by fitting the time-dependence to a linear function at larger times and extrapolating into the cross-talk region, as shown in the left panel of Fig.~\ref{fig_dt}.  The fit was performed for time differences greater than 100~ns, and the integral of the fit was used to calculate the expected signal in the absence of cross-talk.  A similar analysis was performed for next-nearest channels.  The ratio of actual counts to extrapolated counts for adjacent and next-nearest channels is shown in the right panel Fig.~\ref{fig_dt}.  Note that for $\Delta{\rm ch}=2$ the correction returns to unity for times that are smaller than the transit time between the next-nearest neighbor channels.  Note that this correction was only used for time differences less than 100~ns, and was not used in the regions above where deviations from unity are dominated by statistical fluctuations.  A cross-talk correction was also calculated for the horizontal neighbors separated by 2~meters ($\Delta{\rm ch}=8$) and was found to be negligible.

The $Y_{2F}$ values corrected for cross-talk are listed on the second row of each time-gate in Table~\ref{tab_y2f_vstgate} and plotted as solid symbols in Fig.~\ref{fig_y2f_vstgate}.  Note that the corrected $Y_{2F}$ values for beryllium are consistent with zero as expected.  The difference between $Y_{2F}$ values for the DU and HEU stacked discs, already visible before the application of the correction, is still easily discernable in Fig.~\ref{fig_y2f_vstgate}.  Although the $Y_{2F}$ difference grows larger with increasing time-gate, the most significant difference for this measurement, approaching 5-sigma, occurs in the region of 50--100~ns.  A time-gate of 100~ns is applied to the analysis of all subsequent measurements.

\section{Results and Simulation Benchmarks}
\label{sec:results}

\begin{figure}
\begin{center}
\includegraphics[width=0.8\textwidth]{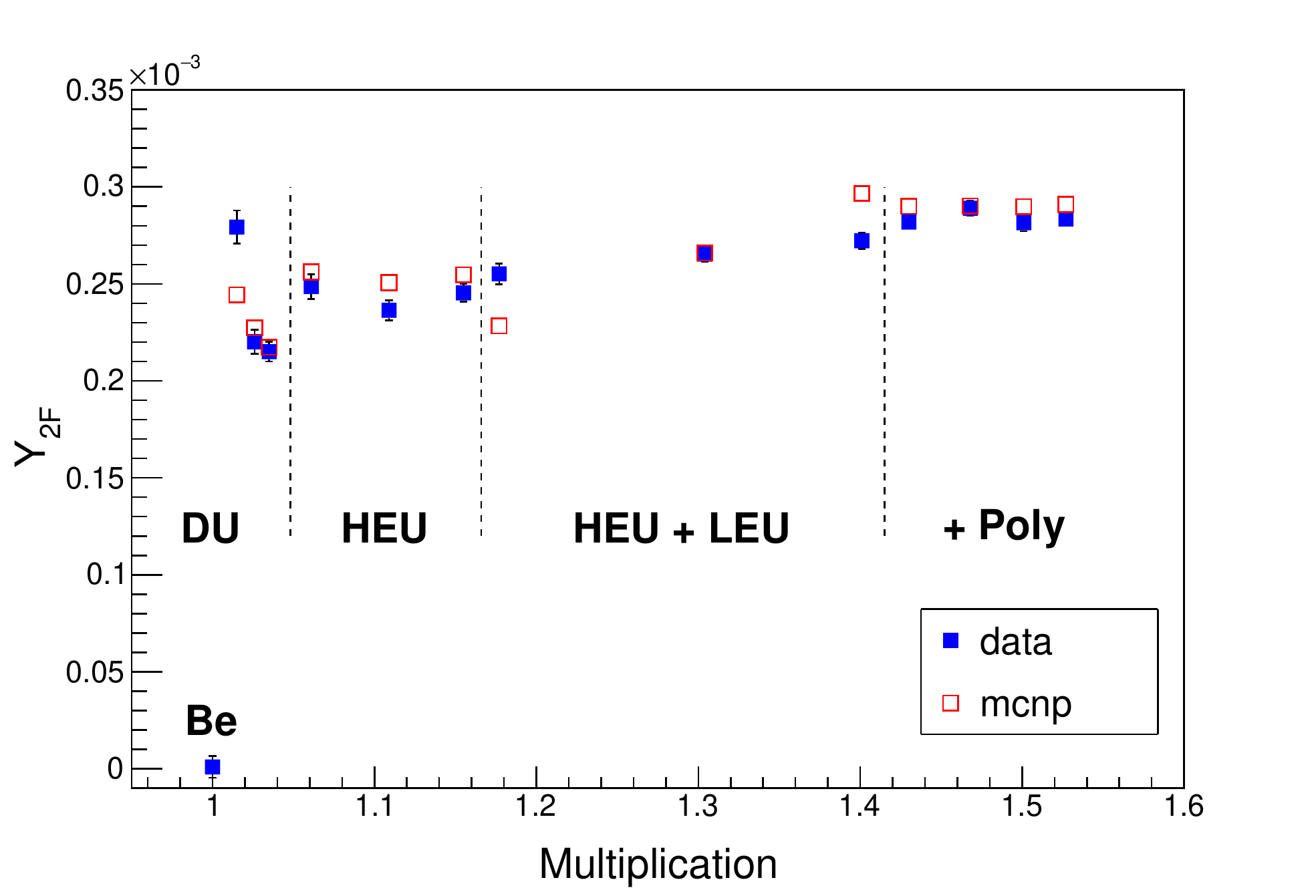}
\end{center}
\caption{Measured $Y_{2F}$ for 100 ns time-bins {\em vs.} Multiplication for all SNM objects measured at the PSI facility in March, 2013.  Corresponding MCNP calculations are plotted as open squares.}
\label{fig_y2f_data_mcnp}
\end{figure}

The $Y_{2F}$ values for 100 ns time-gates and cross-talk corrections for the complete set of objects listed in Table~\ref{tab_objects} are plotted in Fig.~\ref{fig_y2f_data_mcnp} as a function of object multiplication calculated with MCNP.  Contrary to expectations based on Eq.~\ref{eq:y2f_induced}, the $Y_{2F}$ values do not depend simply on the multiplication.  The $Y_{2F}$ values decrease with multiplication for the DU discs, remain approximately constant for the HEU disks, and finally exhibit the expected monotonic increase for the composite systems.  These results are compared to MCNP simulations of the PSI beam incident on the fissile objects.
The detector acceptance was approximated by applying a $\pm$15-degree angular cut above and below the horizontal.
The angular cut matches the angle subtended by the PNPF detector array, and it is required to account for the polar-angle dependence of the neutron emission rates, which varies with object and is strongest for the LEU+HEU+HDPE composite objects.
A minimum energy cut of 1.5~MeV is also applied to the neutrons to approximate the online energy cut implemented in the DAQ.  Neutrons within this acceptance region were used to calculated the $Y_{2F}$ in the same manner as the data, using 100~ns time-gates, but without the need for cross-talk corrections.  A geometric efficiency factor of 0.25\% was applied to the simulated $Y_{2F}$ values.  This value was determined empirically and is comparable to the 0.243\% geometric detector efficiency calculated in GEANT4.  The MCNP results reproduce the overall trends of the data, although the simulations under-predict the decrease for DU and over-predict the increase for the LEU+HEU composite object.  The highest multiplication objects consisting of the interleaved LEU and HEU objects with HDPE moderation are reasonably well-reproduced by the MCNP simulations.  %

This complex dependence of the $Y_{2F}$ values that was observed in the data and qualitatively reproduced by the simulation led us to reconsider some of the assumptions within Eq.~\ref{eq:y2f_induced}.  One important modification is to account for the fact that in photon-induced fissions, the initial fission differs from those of subsequent neutron-induced fissions in the chain.  This is analogous to the modification required for spontaneous fission-chains developed by Prasad and Snyderman~\cite{Prasad2012}.  
The initial photo-fission is especially important for small multiplication values, where the neutrons from the initial fission form a non-negligible contribution to $Y_{2F}$.  We introduce $\nu_\gamma$ for the mean number of neutrons produced in a photo-fission, and $\nu_{\gamma 2}$ for the half-variance.  We also account for the non-fission photo-production and absorption in the extended object by adding the terms $f_o$ for the fraction of neutrons emitted from fission over the total number of neutrons produced by all nuclear reactions and $\epsilon_o$ for the neutron absorption within the object.  The neutron detection efficiency is denoted separately as $\epsilon_d$.  Accounting for these additional factors yields the following expression for $Y_{2F}$ for photon-induced fissions in an extended object,
\begin{equation}
Y_{2F}=f_o \epsilon_o \epsilon_d \left[\frac{\nu_{\gamma 2}}{\nu_{\gamma}}+ \frac{\nu_2}{\nu}(M-1)\right]\left[1+\frac{\nu-1}{\nu}(M-1)\right]
\label{eq:y2f_full}
\end{equation}

\begin{figure}[h]
\begin{center}
\includegraphics[width=0.8\textwidth]{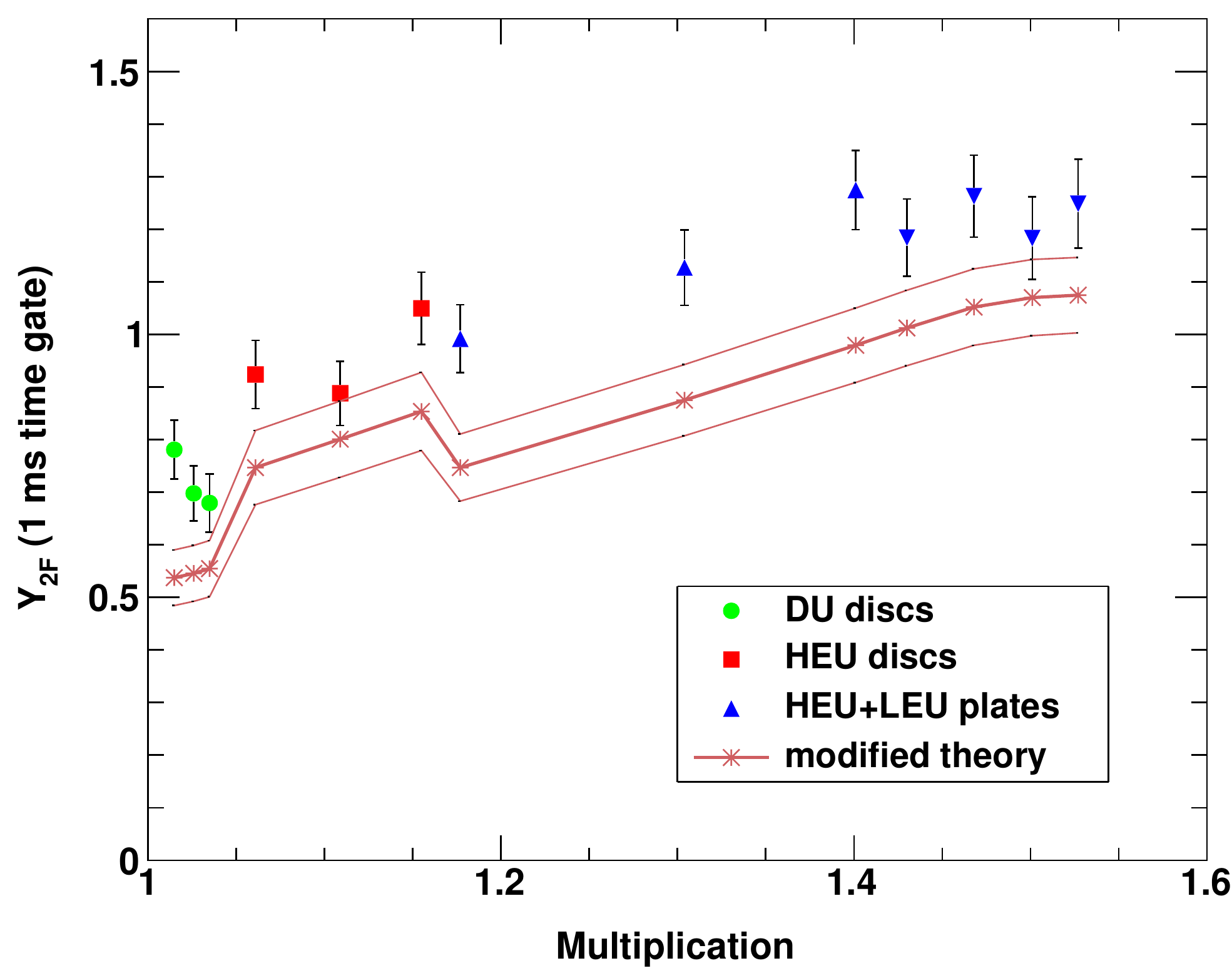}
\end{center}
\caption{Comparison of Eq.~\ref{eq:y2f_full} to MCNP simulations of $Y_{2F}$ {\em vs.} Multiplication for all SNM objects measured at the Passport facility.  The lines representing Eq.~\ref{eq:y2f_full} show the impact of a 10\% variation in the value of the photo-fission neutron variance, $\nu_{\gamma 2}$, which is currently unmeasured.}
\label{fig_y2f_mcnp_theory}
\end{figure}

The mean number of neutrons per photon-fission, $\nu_g$ has been measured by Caldwell~{\it et al.}~\cite{Caldwell1980} and evaluated by  Chadwick~\cite{Chadwick2006}, however, the value for $\nu_{\gamma 2}$ has not yet been measured.  To estimate this value within the simulation, a separate simulation of the MCNP photo-fission package was performed using the PSI energy spectrum as input.  The mean and half-variance values for photon-induced fissions obtained in this way are listed in Table~3.  We also list the corresponding values for neutron-induced fissions as reported by Zucker and Holden~\cite{Zucker1986}.
\begin{table}
\begin{center}
\begin{tabular}{|l|c|c|c|c|}
\hline
Enrichment & $\overline{\nu}$ & $\overline{\nu_2}$ & $\overline{\nu_{\gamma}}$ &$\overline{\nu_{\gamma 2}}$ \\
\hline
DU (0\%)          & 2.43 & 2.45 & 2.78 & 3.09 \\
LEU (20\%)      & 2.45 & 2.52 & 2.75 & 3.04 \\
HEU (93\%)     & 2.51 & 2.63& 2.63 & 2.90 \\
U-235 (100\%) & 2.52 & 2.65 & 2.62 & 2.86 \\
\hline
\end{tabular}
\caption{First and second combinatorial moments for photon and neutron induced fission for various levels of uranium enrichment.  Neutron induced moments are calculated from Holden and Zucker tables for 1--2~MeV neutrons reported in the documentation for the MCNP fission package.  Photon induced moments are calculated from the photo-fission model using the PSI beam energy spectrum.}
\end{center}
\label{tab_nu}
\end{table}
To compare our simulations to Eq.~\ref{eq:y2f_full}, which does not include energy dependence, we remove all energy cuts from the analysis, and we lengthen the $Y_{2F}$ time-gate for this simulation analysis to 1~ms, to encompass the correlation-time for all neutrons, including thermal.  The values for $f_{o}$ and $\epsilon_{o}$ were taken from the simulations and a perfect detector efficiency ($\epsilon_{d}=1$) was assumed.  The full comparison between this MCNP simulation and Eq.~\ref{eq:y2f_full} is shown in Fig.~\ref{fig_y2f_mcnp_theory}.
Eq.~\ref{eq:y2f_full} accounts for the full multiplication dependence observed in the simulations, but consistently under-predicts the $Y_{2F}$ values by 20\%.  One possible explanation for this discrepancy is the lack of energy dependent terms in Eq.~\ref{eq:y2f_full}.  It is also possible that our estimate for the unmeasured $\nu_{\gamma 2}$ differs from the value used in the full simulation.  The $Y_{2F}$ values can be very sensitive to the value of $\nu_{\gamma 2}$ for objects with low multiplication.  To illustrate the sensitivity of $Y_{2F}$ to the value of $\nu_{\gamma 2}$, we vary its value by $\pm 10\%$, as shown by the upper and lower bands.

\section{System Performance Predictions}

With the MCNP simulations benchmarked by experimental results and qualitatively supported by an underlying theory, additional simulations of shorter exposures with larger masses were performed to predict PNPF system performance for distinguishing between DU and HEU in shielded configurations.  For this study we performed simulations for a set of 24 solid spheres of DU and HEU spanning a range of masses from 10 grams up to 50 kilograms.  The masses, radii, and multiplication values are given in Table~4.  MCNP simulations were performed for these objects using a PSI beam of 200~$\mu$A beam for a 10~second exposure.  The emitted neutrons were collected over the entire acceptance and multiplied by the overall geometric efficiency factor of 0.25\% to match the experiment.  A lower minimum energy cut of E$>$1MeV was used, based upon potential improvements to the neutron identification, and a nominal time-gate of 100~ns was used for calculated $Y_{2F}$ values.
\begin{table}
\begin{center}
\begin{tabular}{ c c c c }
\hline
Mass (kg) & radius (cm)    & M-DU   & M-HEU \\ \hline \hline
0.01  &  0.500  &  1.016   & 1.064   \\ 
0.02  &  0.630  &  1.020   & 1.082   \\ 
0.05  &  0.855  &  1.027   & 1.115   \\ 
0.1    &  1.077  &  1.033  & 1.149    \\ 
0.2    &  1.357  &  1.041  & 1.197    \\ 
0.5    &  1.842  &  1.056  & 1.289    \\ 
1       &  2.321  &  1.071  & 1.398    \\
2       &  2.924  &  1.088  & 1.554    \\
5       &  3.969  &  1.115  & 1.941    \\
10     &  5.000  &  1.143  & 2.545    \\
20     &  6.300  &  1.173  & 4.107    \\
50     &  8.550  &  1.219  & 104.1    \\
\hline
\end{tabular}
\caption{Mass, radius, and Multiplication values for DU and HEU spherical objects interrogated with 9-MeV {\it bremsstrahlung} gamma distribution in simulation.}
\end{center}
\label{tab_simObjects}
\end{table}

\begin{figure}
\begin{center}
\includegraphics[width=0.80\textwidth]{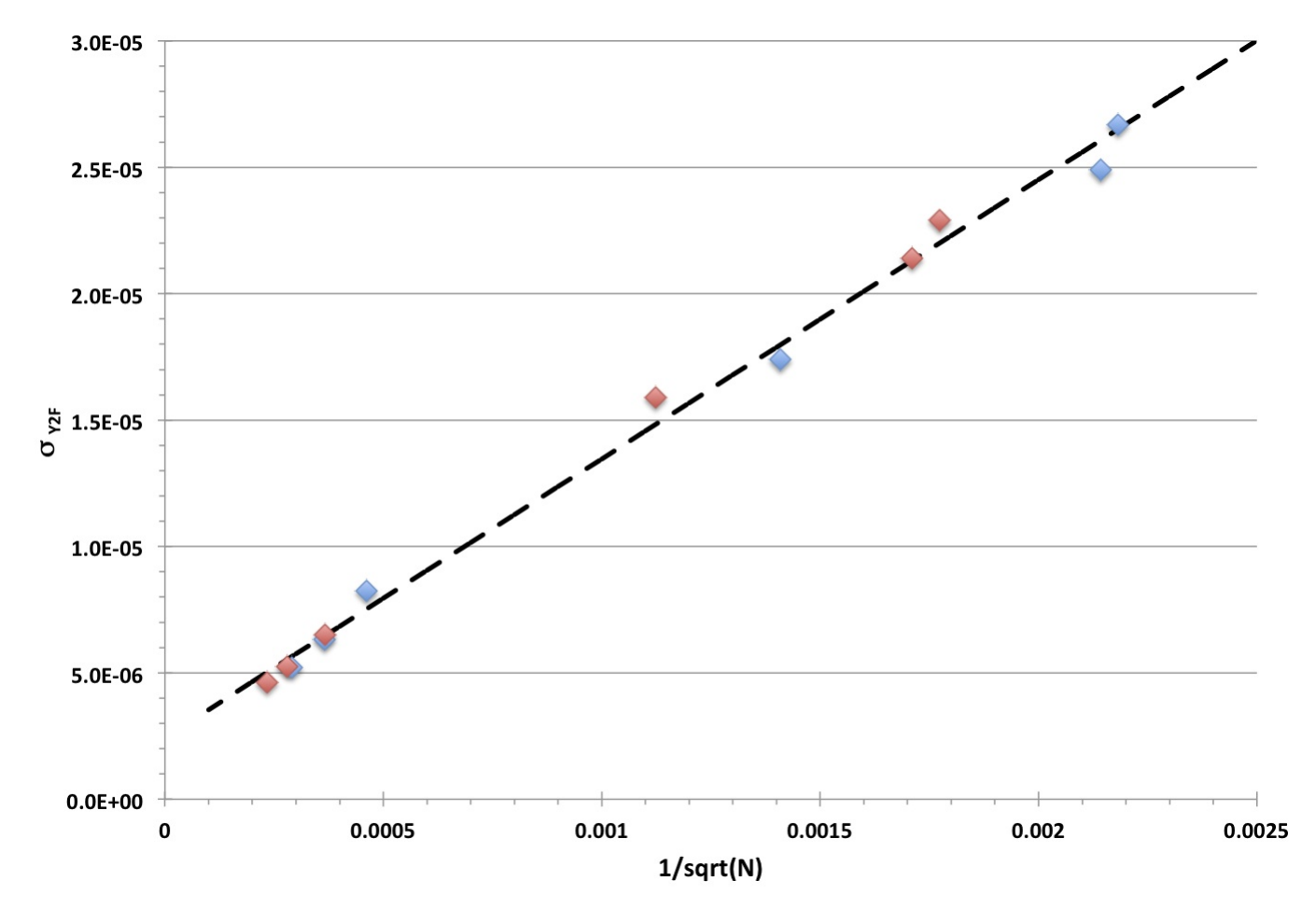}
\end{center}
\caption{Data for the 300s and full duration runs. Red diamonds are the HEU points, blue diamonds the DU points. The uncertainty scales with the inverse square root of the total counts.}
\label{fig_scaling_data}
\end{figure}

Although the $Y_{2F}$ values are calculated directly from the simulations, we chose to use experimental data to extrapolate errors.  Fig.~\ref{fig_scaling_data} shows a uniform, linear dependence in the $Y_{2F}$ error as a function of $N_{counts}^{-1/2}$, independent of object type.  The data are well fit to the following functional form,
\begin{equation}
\sigma_{Y_{2F}}(N_{\mbox{\tiny counts}}) = 2.43\times 10^{-6} + \frac{0.011}{\sqrt{N_{\mbox{\tiny counts}}}}.
\label{eq_sigmay2f_scale}
\end{equation}
We note that this formula does not extrapolate to zero error for zero counts, consistent with the fact that the errors are not strictly Poisson-distributed.  This scaling is used to rescale our simulations for different objects and shielding configurations to predict the dependence on exposure time.

Using the simulated $Y_{2F}$ values and errors extrapolated from data, we use a Gaussian classifier to establish Receiver Operating Characteristics (ROC) curves for the purpose of distinguishing between HEU and DU objects.  Using the HEU as the alarm scenario and the DU as the null scenario we can assign as ``positive'' value based on a Gaussian probability threshold:
\begin{equation}
\mathcal{P}=1 - \int^\infty_T dY_{2F} \frac{1}{\sqrt{2 \pi}\sigma}e^{-(Y^{exp} - Y_{2F})^2/2\sigma^2}
\end{equation}
where $Y^{exp}$ is the measured $Y_{2F}$ and $\sigma$ is the extrapolated error.

\begin{figure}[th]
\begin{center}
\includegraphics[width=0.95\textwidth]{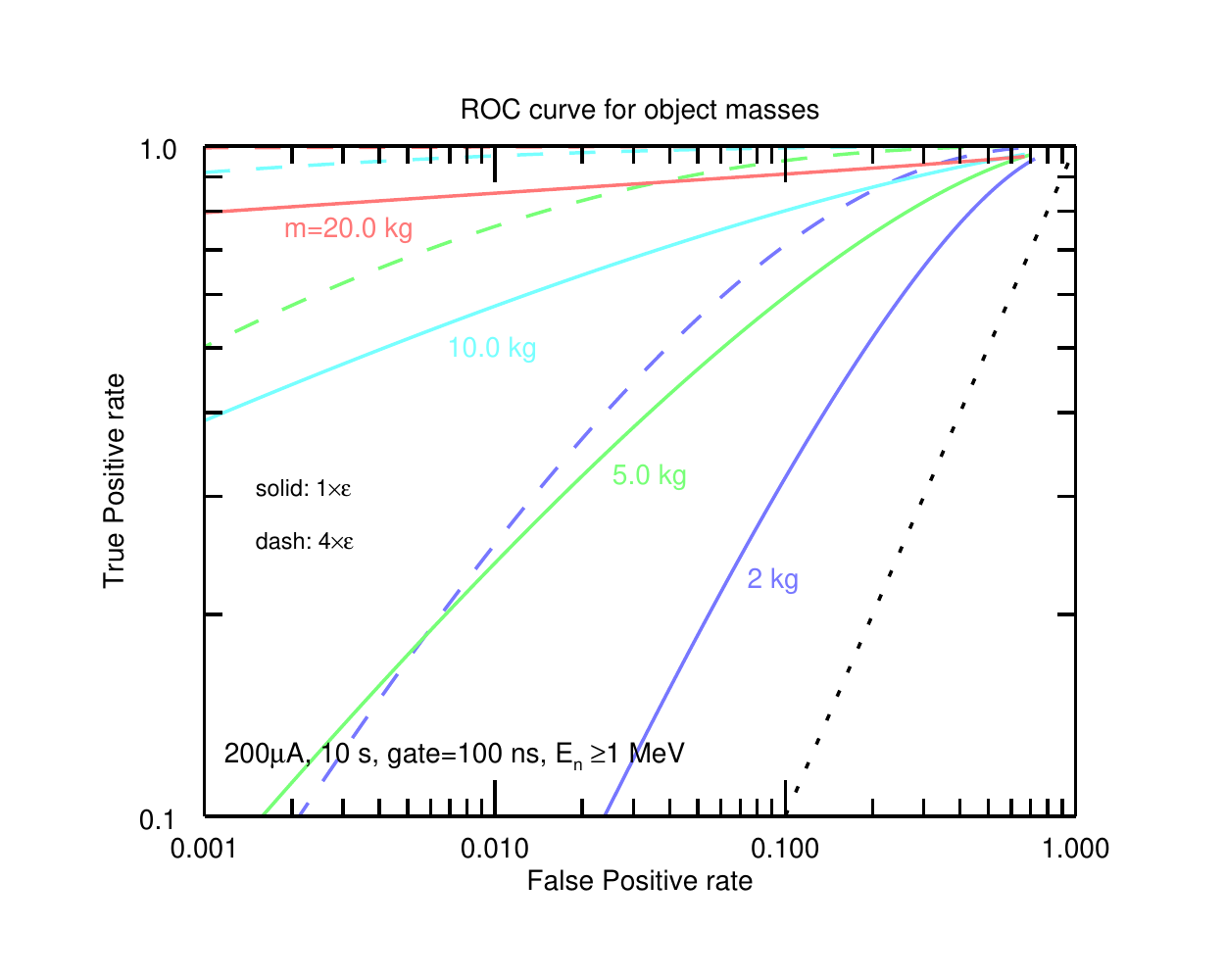}
\caption{ROC curves for distinguishing HEU from DU in 10~seconds of 200~$\mu$A beam for 2, 5, 10, and 20~kg spherical objects.  Solid lines are for the existing Passport system efficiency, dashed lines show the improvement for a $4\times$ increase in efficiency.  The curve for 20kg object at $4\times$ efficiency is consistent with True Positive Rate = 1.} 
\label{fig_ROC_object_mass2}
\end{center}
\end{figure}

The ROC curves for different unshielded object masses are shown in Fig.~\ref{fig_ROC_object_mass2} for spherical objects of 2, 5, 10, and 20~kg. For every object the ROC values were determined for the current PSI detector configuration (solid lines) and for one with $4\times$ increased geometric efficiency.  The analysis shows that while detection of small objects, e.g. 2kg, may be difficult, for larger objects the signal becomes large enough for a fast detection.  This becomes especially true for a larger detector array.  Note that the positive detection of 20~kg sphere with the $4\times$ augmented detector array approaches 100\% with a very low rate for false positives.

\begin{figure}[h]
\begin{center}
\includegraphics[width=0.49\textwidth]{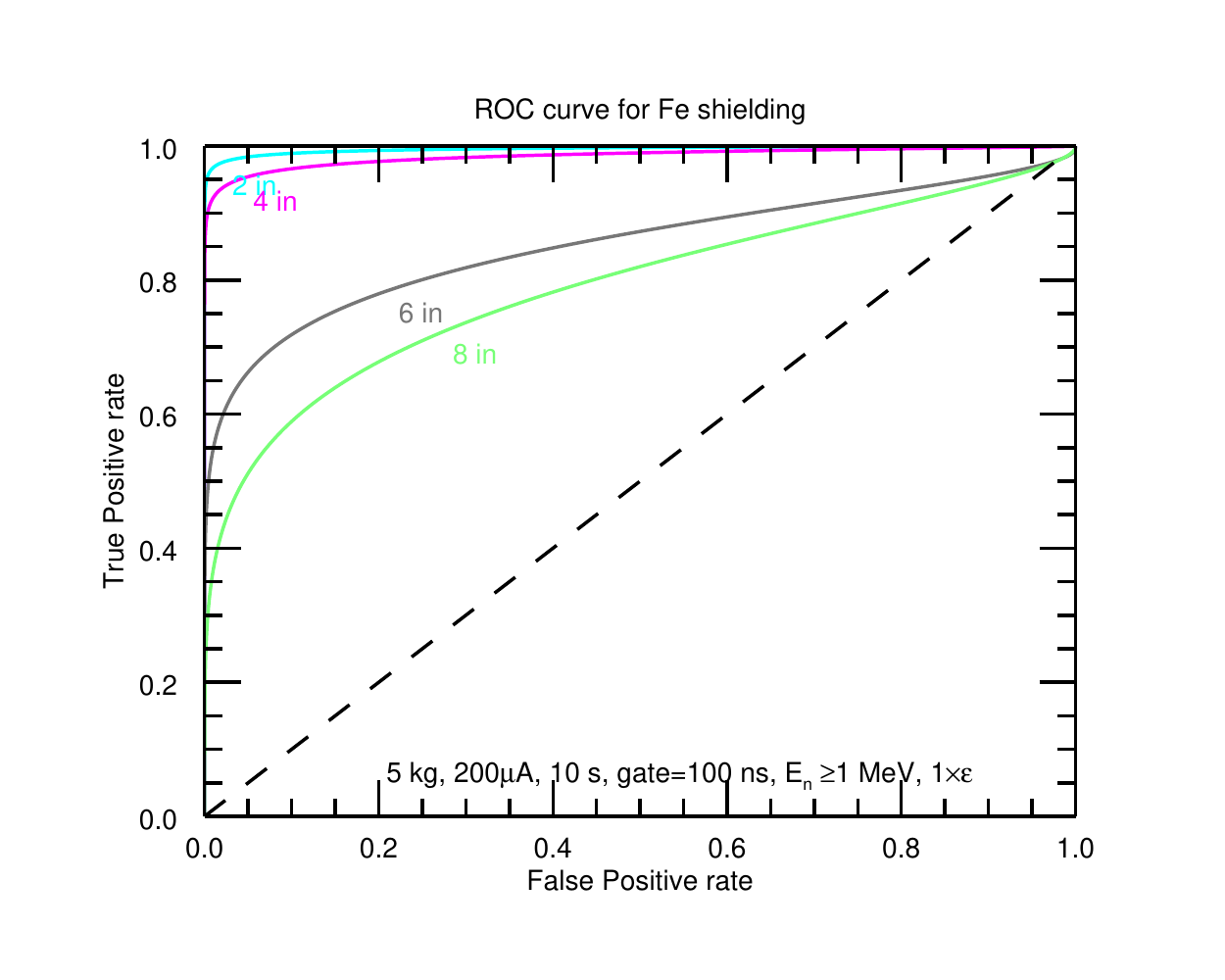}
\includegraphics[width=0.49\textwidth]{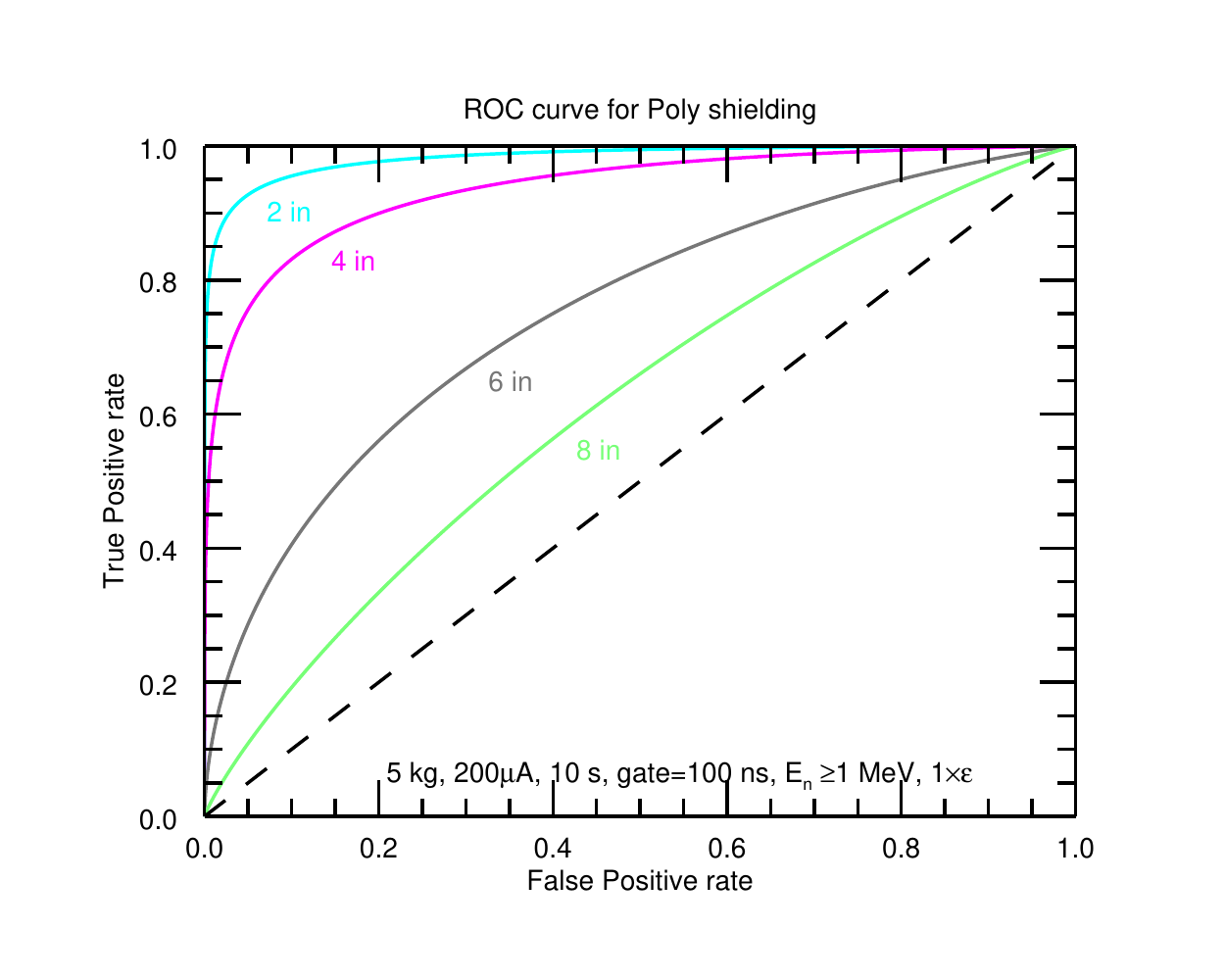}
\caption{ROC curves for distinguishing HEU from DU with 10~s of 200~$\mu$A beam for 5~kg objects with 2, 4, 6, 8 inches shielding of Fe (left) and HDPE (right) spherical shells surrounding the objects.}
\label{fig_ROC_object_shielding}
\end{center}
\end{figure}

In addition to characterizing performance for unshielded objects, the system was also studied for shielded configurations, where the 5~kg objects were surrounded by iron and HDPE blocks of varying thickness.  The first shielding scenario has the primary effect of reducing the incident photon flux, thus reducing statistics in a fixed measurement time.  The second scenario's impact primarily consists of moderating the fast neutrons, and thus reduces neutron statistics.  Figures~\ref{fig_ROC_object_shielding} shows the ROC curves for the 5~kg object for several thickness of full density iron (left) and HDPE shielding (right). 

\section{Conclusions}

The objective of the experimental effort described in this work was to extend the already existing Prompt Neutrons from Photofission (PNPF) technique, described in Ref.~\cite{pnpf-short}, to distinguish between fissionable (e.g. DU) and fissile (e.g. HEU, WGPu) materials by using neutron multiplicity analysis.  The goal of the initial stages of the experimental program was to develop the necessary statistical and mathematical formalism, the Monte Carlo simulation infrastructure, and to experimentally demonstrate the feasibility of the technique.
The results showed a 5-sigma difference in the signal for the two types of objects, proving the feasibility of this methodology.

The next stages of the program focused on additional experiments, involving more complex target configurations using DU, LEU, and HEU.  The $Y_{2F}$ measurements were compared to simulations, which in turn were used to develop a theoretical-numerical model providing a basis for extrapolations to larger objects and shorter exposure times.  The results showed that while the $Y_{2F}$ signal is marginal for practical differentiation of the small objects used in the experiment, for 5kg spherical sizes the neutron multiplication $M$ is large enough to differentiate fissile material types in shielded cargo configurations with reasonable scan times.  

While both the feasibility and practicality of the methodology was demonstrated, significant additional work remains to be done.  Measurements on a wider range of object sizes, both large and small, would be valuable for further understanding the dynamics of $Y_{2F}$.  Also high statistics measurement of photo-fission neutron distributions with thin-foil U-235 and U-238 targets with large acceptance detector that would enable precise and accurate determination of $\nu_{\gamma 2}$ would significantly reduce uncertainties for future system performance studies.

\section{Acknowledgements}
The authors wish to acknowledge helpful conversations with Dr. George Chapline and Prof. William Bertozzi.  The authors thank Cody Wilson of PSI for his help in running the accelerator and the system.
This work is supported in part by the U.S. Department of Homeland Security Domestic Nuclear Detection Office through 
a collaboration between Lawrence Livermore National Laboratory and Passport Systems, Inc.
Lawrence Livermore National Laboratory is operated by Lawrence Livermore National Security, LLC, for the U.S. Department of Energy, National Nuclear Security Administration under Contract DE-AC52-07NA27344.

\section*{References}

\providecommand{\noopsort}[1]{}\providecommand{\singleletter}[1]{#1}%

\end{document}